%% file: moriond.tex
\documentclass[11pt]{article}
\usepackage{moriond,epsfig}

\def\PLB{{\em Phys. Lett.}  B}

\def\be{\begin{equation}}
\def\ee{\end{equation}}
\def\bea{\begin{eqnarray}}
\def\eea{\end{eqnarray}}

\include{my_definitions}

\begin{document}
\vspace*{4cm}
\title{HIGGS PHYSICS AT THE LARGE HADRON COLLIDER
\footnote{Talk Presented at
the ``XXXVI Rencontres de Moriond Electroweak Interactions and Unified
Theories'', 10-17 March 2001}
}

\author{ Davide Costanzo }

\address{INFN and University of Pisa, \\ 
I--56010 S. Piero a Grado -- Pisa, Italy}

\maketitle\abstracts{A major goal of the future Large Hadron
Collider will be the Higgs boson search. In this paper the discovery
potential is described as a function of the Higgs mass showing that
a Standard Model Higgs boson can be discovered after less than two years 
of running
of the collider. The MSSM Higgs searches and the precision achievable on
the measurement of the Higgs boson parameters are also discussed.}

\section{Introduction}

The main focus of the experiments at the future proton-proton Large
Hadron Collider (LHC) at CERN will be the investigation of the nature
of the electroweak symmetry breaking, and therefore the search for the
Higgs boson.
Two general purpose experiments, ATLAS\cite{atlas} and CMS\cite{cms}, have
been designed to study collisions at the LHC, and have been optimized to
cover a large spectrum of possible physics signature, accessible at the
high luminosity and center-of-mass energy of the (LHC).

\begin{figure}
\begin{center}
\psfig{figure=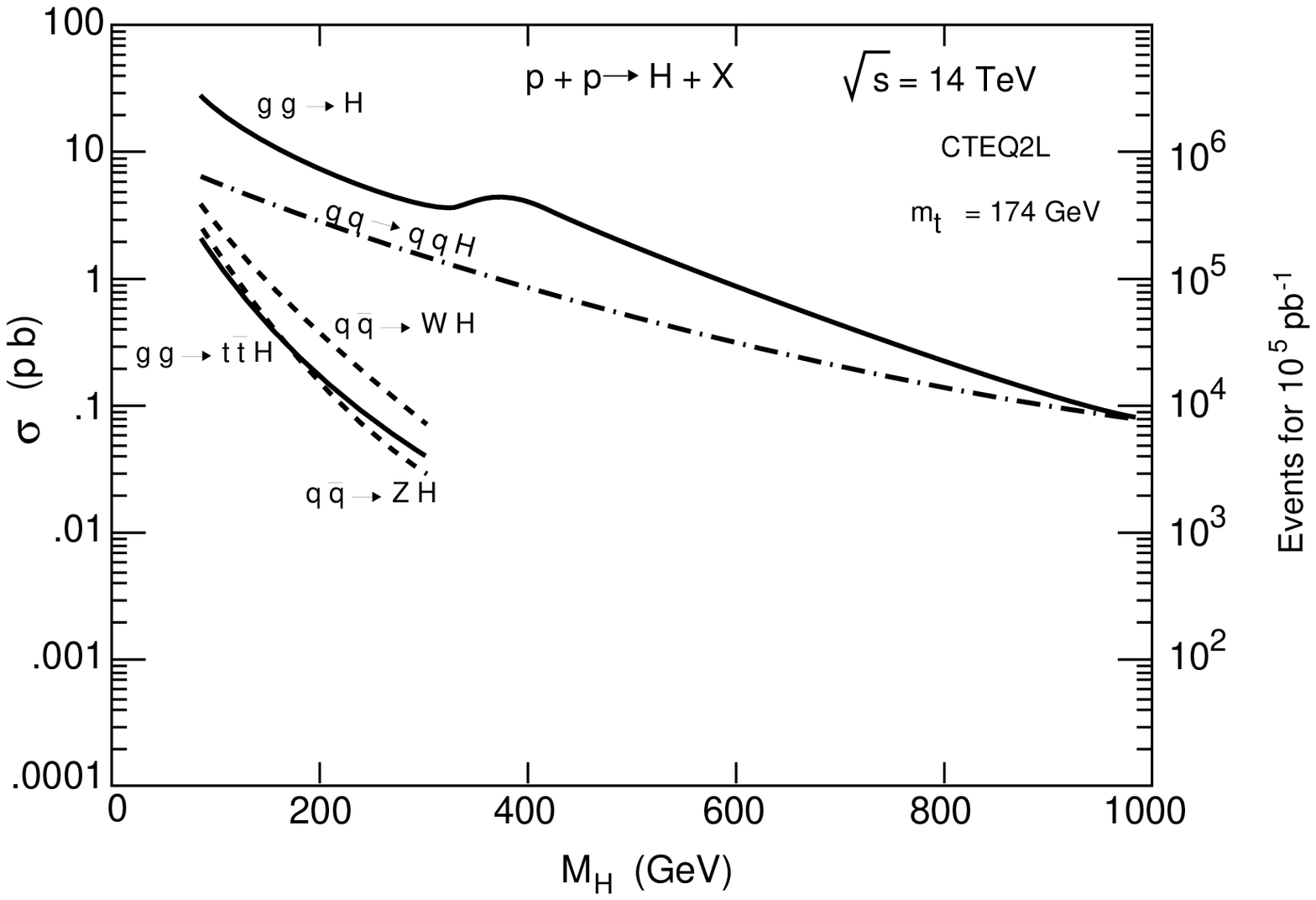,height=2.in}
\psfig{figure=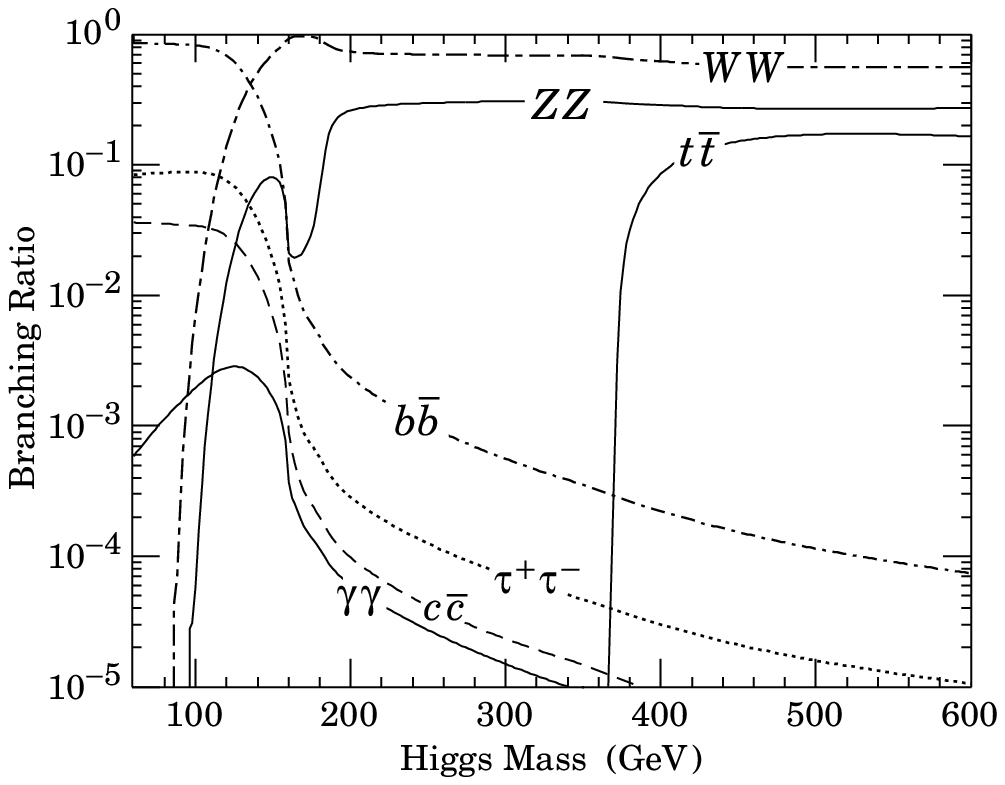,height=2.in}
\end{center}
\caption{Higgs production cross section at the LHC  and Higgs
branching ratios as a function of $m_H$.
\label{fig:higgscross}}
\end{figure}

Detailed simulations of both experiments have been performed to
demonstrate that the Higgs should be discovered regardless of its mass
which is an unknown parameter of the Standard Model. The Higgs can be
produced at the LHC through several mechanisms, whose individual cross
section is plotted in Fig.~\ref{fig:higgscross} (left) as a function of
the Higgs mass ($m_H$)\cite{spira}. It can be seen that the inclusive
production
dominates over the entire mass range, followed by $Hqq$, while
the production in association with a $W$, $Z$, or $\Ttbar$ pair has a
sizeable contribution for $m_H \leq 200$~GeV. The branching fractions of
the Higgs decays vary as a function of $m_H$ as shown in
Fig.~\ref{fig:higgscross} (right)\cite{spira}. 
%
%The physics potential of the ATLAS experiment in this area is briefly
%summarized in the following. For details the reader is referred to the
%{\it ATLAS Detector and Physics Performances Techical Design Report},
%where all relevant ATLAS studies on the physics potentials are
%documented.
%
\section{LHC running scenarios and cross sections}
Throughout the paper, it is assumed that an initial luminosity of
$10^{33} \mbox{cm}^{-2} \mbox{s}^{-1}$ (hereafter called {\it low
luminosity}) can be achieved at the LHC. It is expected that this value
should rise,
during the first two years of operation, to the design luminosity of
$10^{34} \mbox{cm}^{-2} \mbox{s}^{-1}$ (hereafter called {\it high
luminosity}). Integrated luminosities of about 10~fb$^{-1}$, 30~fb$^{-1}$,
100~fb$^{-1}$, 300~fb$^{-1}$ should therefore be collected after one year,
three years, four years and less than ten years of data taking,
respectively.

The physics performance has been evaluated using the following
assumptions: (i) Despite the considerable progress in the calculation of
higher order QCD corrections over the last years, these corrections
(K-factors) are not known for all signal and background processes of
interest. Therefore, the studies presented here, have consistently and
conservatively refrained from using K-factors, resorting to Born-level
predictions for both signal and backgrounds. (ii) The non diffractive
inelastic cross section has been assumed to be 70~mb. At low
(high) luminosity,
this leads on average to a superposition of 2.3 (23) minumum bias events
per
crossing on top
of the hard collision. These so called pile-up contribution have been
included for both low and high luminosity. (iii) Physics process have been
simulated with the Pythia Monte Carlo program\cite{pythia}, including
initial- and
final-state radiation, hadronization and decays. Although many results
have
been obtained using a fast simulation of the detectors, all key
performance characteristics have been evaluated with a full GEANT
simulation, both at low and high luminosity. 

\section{The search for the Higgs boson}
The Standard Model Higgs boson is searched for at the LHC in various decay
channels, the choice of which is given by the signal rates and the
signal-to-background ratios in the various mass regions. The search
strategies and background rejection methods have been estabilished through
many studies over the past years\cite{atlas,cms}.

\begin{figure}
\begin{center}
\psfig{figure=sens30.epsi,height=2.3in} ~~~~~~~~~~
\psfig{figure=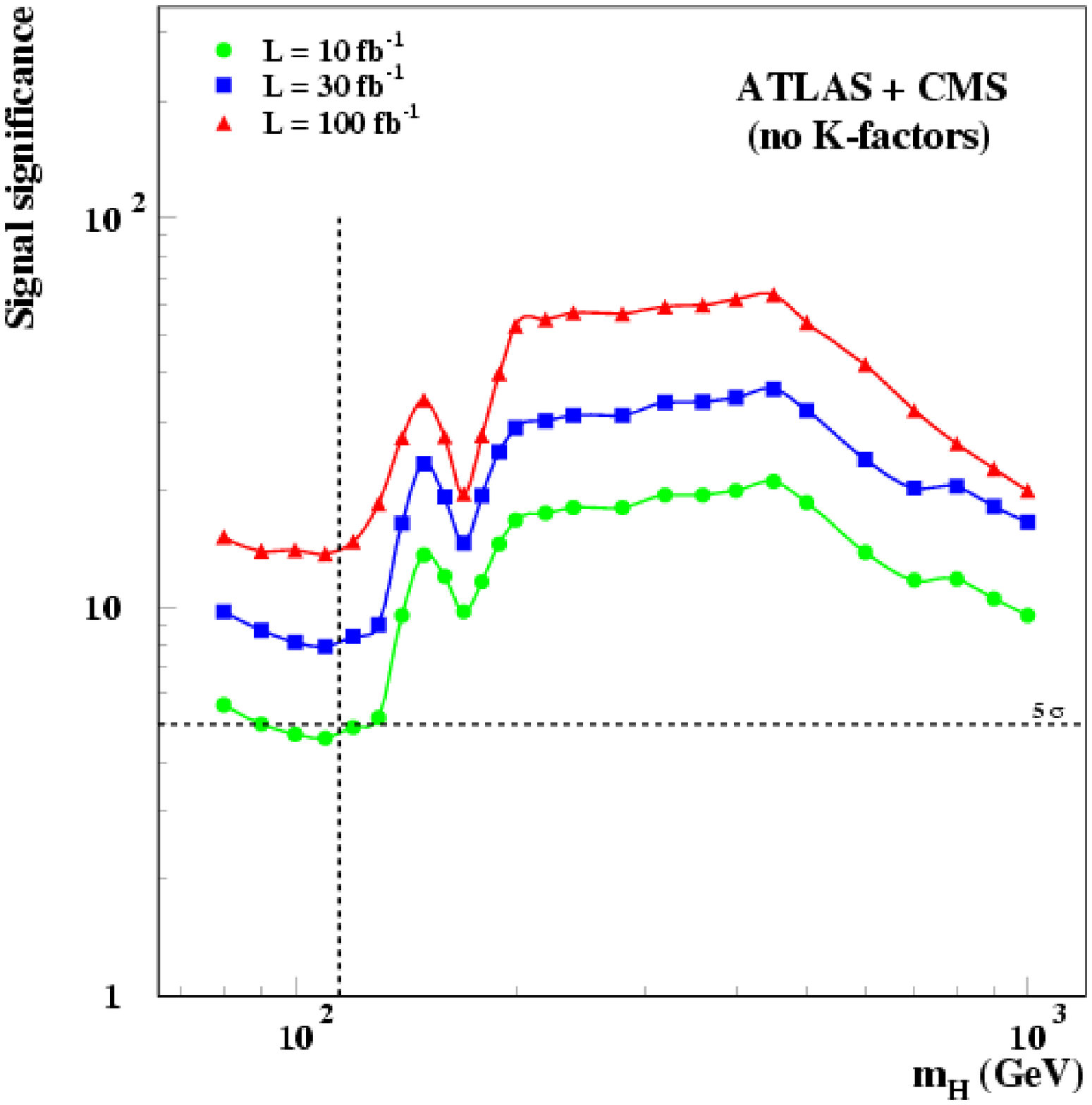,height=2.5in} 
\end{center}
\begin{flushleft}
\caption{Sensitivity for the discovey of a Standard Model Higgs boson as a
function of the Higgs boson mass. The statistical significances are
plotted for individual channels, assuming integrated luminosity of
30~$\Ifb$ (left) for the ATLAS experiment. The overall statistical
significance is plotted (right) as a function of the Higgs mass for three
different integrated luminosity scenarios: 10, 30, and 100~$\Ifb$ for
ATLAS and CMS combined.
\label{fig:sensitivity}} 
\end{flushleft}
\end{figure}

The overall sensitivity for the discovery of a Standard Model Higgs boson
over the mass range from 80~GeV to 1~TeV is shown in
Fig.~\ref{fig:sensitivity} (left). This sensitivity is given for
individual
channels as well as for the combination of the various channels, assuming
integrated luminosities of 30~fb$^{-1}$ for the ATLAS 
experiment. The combined sensitivity for both ATLAS and CMS is shown in
Fig.~\ref{fig:sensitivity} (right) for 10~fb$^{-1}$, 30~fb$^{-1}$ and
100~fb$^{-1}$. A Standard Model Higgs boson can be discovered at the LHC
over the full mass range from the LEP2 lower limit up to the TeV range
with a high significance. A $5\sigma$ discovery could already be achieved
over the full mass range after the first two years of running at low
luminosity. Over a large fraction of the mass range the discovery of a
Standard Model Higgs boson will be possible in two or more independent
channels.

%~~
The main channels which will be used at the LHC to look for a Standard
Model Higgs boson can be classified depending on $m_H$ as 
follow:
\begin{itemize}
\item Low mass region ($m_H < 130$~GeV). 
The search of the Standard Model Higgs boson is challenging in this mass
region. Even though its natural width is narrow, the backgrounds from
$\Ttbar$, $\gamma \gamma$, and W/Z+jets production are relatively large
and thus,
an excellent detector performance in terms of energy resolution and
background rejection is required. Two decays modes are
experimentally important in this region: $H \rightarrow \Bbbar$ and $H
\rightarrow \gamma \gamma$. The first one has a branching ratio close to
100\% in most of this region, and therefore inclusive Higgs production
followed by $H \rightarrow \Bbbar$ has a large cross-section ($\approx$ 20
pb). However, since the signal-to-background ratio for the inclusive
production is smaller than $10^{-5}$, it will be impossible to observe
this channel above the QCD background and even to select it at the trigger
level. On the other hand, the associated production $\Ttbar$H, WH, ZH,
with $H \rightarrow \Bbbar$ and with an additional lepton coming from the
decay of the accompanying particles, has a much smaller cross section ($<$
1 pb) but gives rise to final states which can be extracted from the
background. \\
The $H \rightarrow \gamma \gamma$ channel has a branching ratio at the
level of $10^{-3}$ and therefore a small cross-section ($\approx$
50~fb). However, the signal-to-background ratio ($\approx 10^{-2}$) is
much more favorable than for the inclusive $\Bbbar$ channel. 

In addition it has been recently suggested\cite{zeppenfeld} that the
statistical
significance can be increased by searching for a Higgs boson  produced
through vector boson fusion, e.g. with two forward emitted jets
accompanying it in the final state. The detection of those two energetic
jets provides a powerful tool for background rejection.
 \item Intermediate mass region (130~GeV $< m_H <$ 2 $m_Z$). 
The most
promising channels for the experimental searches are $H \rightarrow
ZZ^{(*)} \rightarrow
4l$ and $H \rightarrow
WW^{(*)} \rightarrow l \nu l \nu$, the latter can be studied both using
inclusive and associated W production.
\item High mass region ($m_H >$ 2 $m_Z$). 
This is the best region to discover a Higgs boson signal at the LHC, since
the $H \rightarrow ZZ \rightarrow 4l$ channel gives rise to a gold
plated signature, almost background free. For very large masses ($m_H >$
500~GeV) searches for this decay mode will be supplemented by searches for
other channels, such as $H \rightarrow ZZ \rightarrow ll \nu \nu$
and $H \rightarrow WW \rightarrow l \nu jj$ (where $j$ stands for
hadronic jet), which have larger branching ratios and therefore can
compensate for the decrease in the production cross-section.
\end{itemize}

The experimental techniques used to extract a possible signal above the
backgrounds and the LHC discovery potential are illustrated here. 

\subsection{$H \rightarrow \gamma \gamma$}
\begin{figure}
\begin{center}
\psfig{figure=hggatlas.epsi,height=2.5in}  ~~~~~~~~~~~~~
\psfig{figure=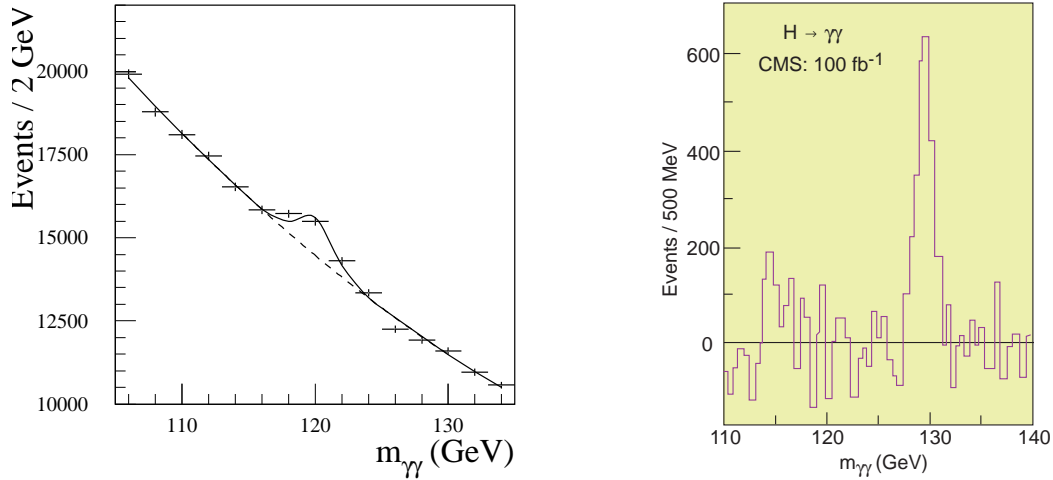,height=2.5in} 
\end{center}
\caption{Expected $H \to \gamma \gamma$ signal at the LHC for an
integrated luminosity of 100~$\Ifb$. Left: the signal recontructed in
ATLAS for $m_H$=120~GeV is shown on top of the irreducible
$\gamma \gamma$ background. Right: the signal reconstructed in CMS for
$m_H$=130~GeV is shown after background subtraction.
\label{fig:hgammagamma}} 
\end{figure}

This channel should allow observation of a Higgs boson over the mass
region up to  150~GeV. The branching ratio for this channel is of
the order of 0.1$\div$0.3\% as the Higgs is coupled to photons only
through a $W$ loop. The final state consists of two high-$p_T$ photons
($P_T \approx 50$~GeV), despite the simple signature, this is the most
challenging channel for the performance of the LHC electromagnetic
calorimeters and has indeed driven the design and the technology choices
for these detectors. The reason is that there are two large backgrounds to
fight. (i) The $\gamma \gamma$ production through QCD diagrams, which is
an irriducible background without a resonant structure, and decreases
smoothly with the invariant mass of the two photons. The $\gamma \gamma$
cross-section is about 60 times larger that the $H \to \gamma \gamma$
cross-section in the region $m_{\gamma \gamma} \simeq 100$~GeV. Therefore,
excellent detector energy and angular resolution are needed in order to
extract a narrow resonant peak above the overwhelming continuum
background.
(ii) The reducible $\gamma$--jet and jet--jet production, where one or
both
jets fake a
photon. This happens when a quark fragments into a very hard $\pi^0$ plus
a few other particles which are too soft to be detected. The two photons
from the subsequent $\pi^0$ decay are very close in space because the
parent $\pi^0$ is usually produced with a large boost. For instance, for a
$\pi^0$ of $p_T \approx 50$~GeV the distance between the two decay photons
is smaller than 1~cm at the front face of the electromagnetic calorimeters
($\approx 150$~cm from the interaction point). 
%Therefore the two photons
%appear as a single photon in the electromagnetic calorimeter, unless the
%latter has a fine enough granularity toallow detection of two distinct
%showers. 
Although the probability for a jet to fragment into a single isolated
$\pi^0$ is small, the cross section for $\gamma$jet and jet--jet
production is $\approx 10^6$ times larger than the cross section for the
$\gamma \gamma$ continuum. Therefore a jet rejection of at least 1000 is
required to suppress this background to well below the irreducible $\gamma
\gamma$ background.
Fig.~\ref{fig:hgammagamma} shows the expected $H \to \gamma \gamma$ signal
in ATLAS above the continuum (reducible and irreducible) $\gamma \gamma$
background, for a Higgs mass of 120~GeV and an integrated luminosity of
100~$\Ifb$. About 1000 events are expected in the Higgs peak. Also shown is
the expected signal in CMS for $m_H=130$~GeV, after background
subtraction. The CMS significance should be about 15\% better than the
ATLAS significance, due to a better energy resolution of the
electromagnetic calorimeter. 
~
\subsection{$H \rightarrow \Bbbar$}
The $H \to \Bbbar$ decay channel can be searched for only if the Higgs
boson is produced in association with other particles. In particular the
$\Ttbar H$ associated production is promising in terms of statistical
significance achievable. The final state is considerably complex, since it
consists of two $W$-bosons and four $b$-jets. For trigger purposes, one of
the $W$ bosons is required to decay leptonically, whereas the other one
is assumed to decay into a $\Qqbar$ pair. In order to reliably extract the
signal, the analysis requires that both top quarks be fully
reconstructed. This method reduces considerably the large combinatorial
background in the signal events themselves, since two of the $b$-jets are
associated to the top decays, and therefore the remaining two should come
from the Higgs boson decay. The signal should appear as a peak in the
$\Bbbar$ invariant mass distribution, above the various background
processes.

\subsection{$H \rightarrow ZZ^{(*)}$}
This channel, which can be observed at the LHC over the mass region
120--700~GeV, gives rise to a very distinctive signature, consisting of
four leptons whose invariant mass is consistent with the nominal Higgs
boson mass.

\begin{figure}
\begin{center}
\psfig{figure=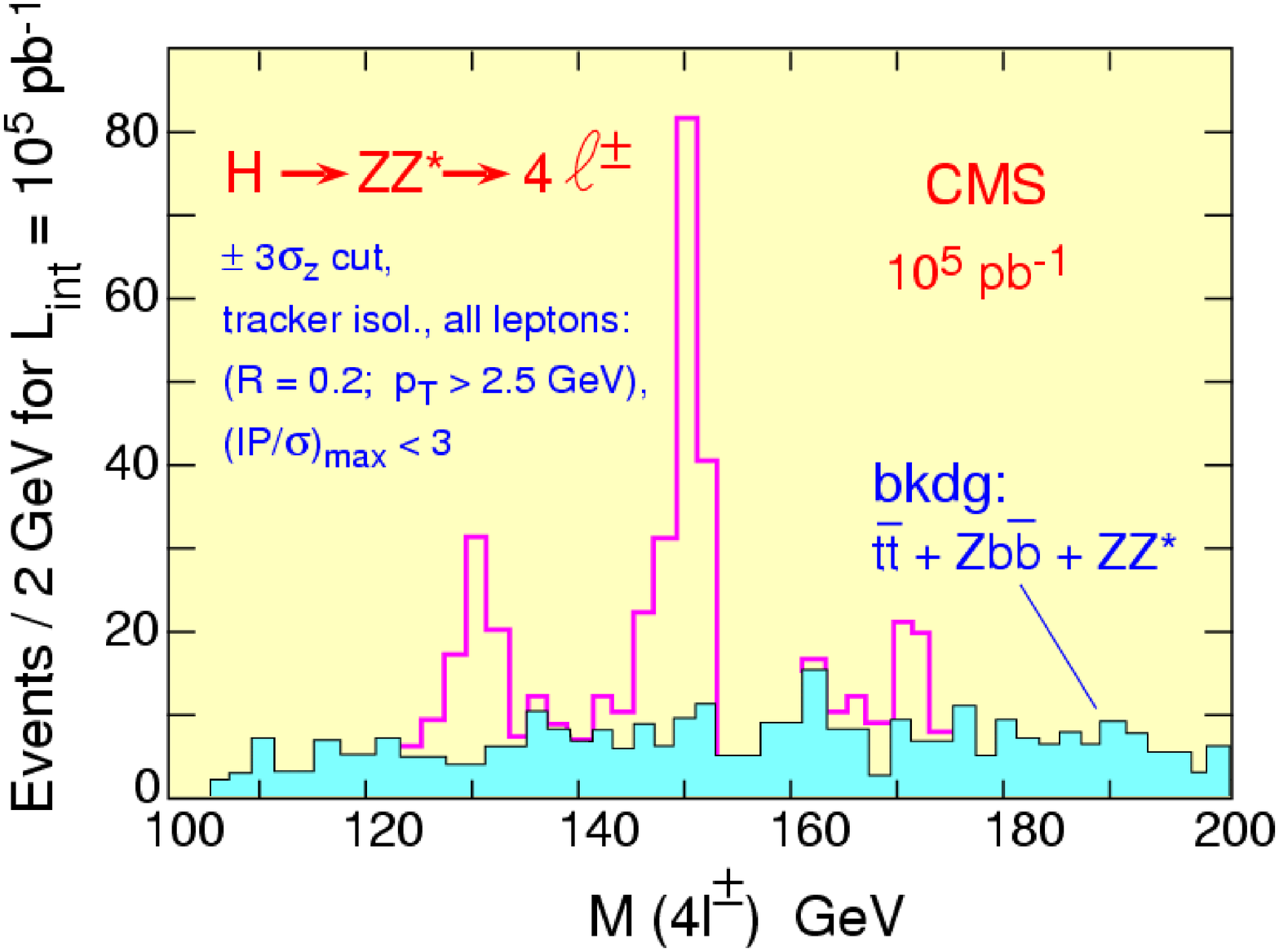,height=2.5in} 
\psfig{figure=hwwint.epsi,height=2.5in} 
\end{center}
\caption{Left: invariant mass distribution of the four leptons for the
process $H \to ZZ \to 4l$. The peaks for $m_H$=130, 150 and 170~GeV are
shown over the SM backgound assiming an integrated luminosity of
100~$\Ifb$ for the CMS experiment. Right: transverse mass for the process
$H \to WW \to l \nu l \nu$ assiming $m_H$=150~GeV and an integrated
luminosity of 30~$\Ifb$ for the ATLAS experiment. Dashed line is the total
background contribution and full line is the sum of signal plus
background.
 \label{fig:hww}} \end{figure}

The expected backgrounds, and therefore the search criteria, depend on the
Higgs mass. If $m_H > 2m_Z$, then both $Z$ bosons in the final state are
real and two pairs of leptons with the same flavor and opposite sign
should have an invariant mass compatible with the $Z$ mass. In this
region the backgrounds, such as the irreducible $pp \to ZZ \to 4l$, are
small. 

For $m_H < 2 m_Z$, on the other hand, the backgrounds are large. In
addition to the already mentioned $pp \to ZZ \to 4l$ continuum, there are
two potentially dangerous reducible backgrounds: $\Ttbar \to 4l + X$ and $
Z \Bbbar \to 4l +X$ with two leptons coming from semileptonic decays of
two $b$-quarks. These backgrounds can be rejected by asking the invariant
mass of at least one lepton pair to be compatible with the $Z$ mass, by
requiring that all leptons be isolated (leptons from the $b \to lX$ decays
are usually non-isolated), and by requiring that all leptons come from the
interaction vertex (leptons from $b$-quark decay are produced at $\approx$
1~mm from the vertex due to the long $B$-hadrons lifetime). 

The signal significance expected in each experiment varies between three
and 25 for an integrated luminosity of 30~$\Ifb$, over the Higgs mass
range 120--700~GeV. For larger masses, the production cross-section
decreases fast and the width increases, and the branching ratio into four
lepton final states is too small to allow observation of this channel. 
Other channels should be used for $m_H>700$~GeV, such as $H \to ZZ \to ll
\nu \nu$ or $H \to WW \to l \nu jj$ which have a larger branching
fractions.

\subsection{$H \rightarrow WW^{(*)}$}
For Higgs boson masses close to 170~GeV, the signal significance in the $H
\to ZZ \to 4l$ channel is reduced, due to the suppression of the $ZZ^*$
branching ratio as the $WW$ decay mode opens up.
% with a branching fraction
%which is about 100 times bigger than the $ZZ$ one (see
%Fig.~\ref{fig:higgscross}). 
The best candidate for Higgs boson
discovery in the mass range 150--190~GeV is the channel $H \to WW^{(*)}
\to l \nu l \nu$. Due to the presence of two neutrinos in the final state,
it is possible to reconstruct only the Higgs transverse mass, and the
signal can  not be observed as a clear mass peak. The transverse mass
distribution for the signal and for the
backgrounds is shown in Fig.~\ref{fig:hww} for an Higgs mass of 150~GeV,
assuming an integrated luminosity of 30~$\Ifb$.
 The observation of a Higgs signal is based on a
counting experiment, and the systematic on the knowledge of the total
background yield is crucial. A 5\% systematic uncertainty on the total
number of background events is assumed here, and used to derive the
statistical significance for the channel. 

To confirm the discovery one can exploit  production processes  where the
Higgs
boson is produced in
association with a W boson. This channel provides a distinctive signature
with 3 W bosons in the final state and with a  low level 
of background, but with a lower --by a factor of
about 10-- event rate.  Using this associated production channel the
statistical significance is
worse than in the inclusive channel, however  a good signal
to background ratio (of the order of 2) can be achieved so that the
results are not so sensitive to the precise knowledge
of the absolute normalization of the backgrounds.

The final state in which all the $W$ bosons decay into lepton pairs and
that in which two
of them decay leptonically and one hadronically have been studied. In
the latter case, to obtain 
the needed rejection against the $\Ttbar$ production followed by a
di-lepton decay as well as against many other physics backgrounds with
opposite sign leptons in the final state, it is required that the two
charged leptons  have the same sign. 

\section{Supersymmetric Standard Model Higgs}
\begin{figure}
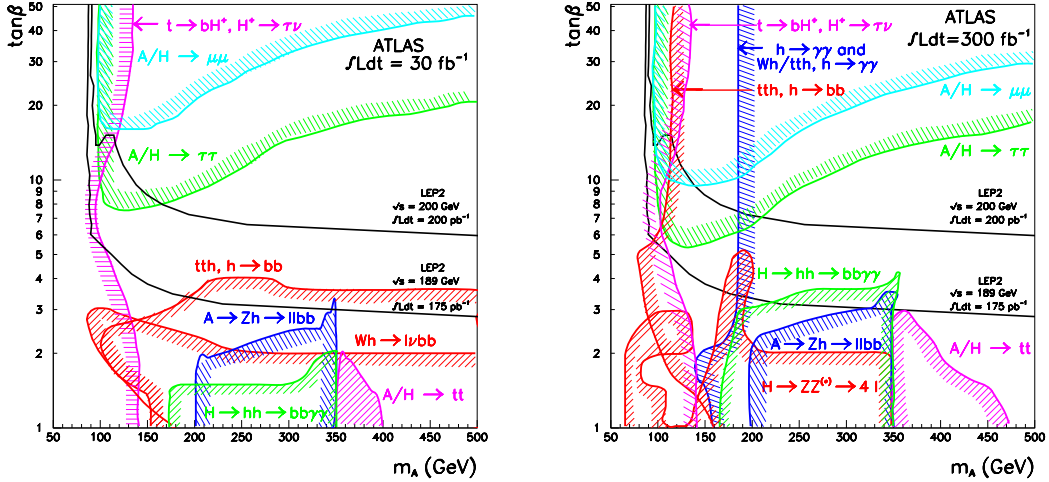

\begin{center}
\psfig{figure=susy2.epsi,height=2.5in} ~~~~~~
\psfig{figure=susy.epsi,height=2.5in} 
\end{center}
\caption{ATLAS sensitivity for the discovery of MSSM Higgs bosons (in the
case of maximal mixing). The $5 \sigma$ discovery contour curves are shown
in the ($m_A$, $\tan \beta$) plane for individual channels for integrated
luminosity of 30~$\Ifb$ (left) and 300~$\Ifb$ (right).  
 \label{fig:susycontour}} \end{figure}

The LHC experiments have also a large potential in the investigation of
the Mimimal Supersymmetric Standard Model (MSSM) Higgs sector. If the SUSY
mass scale is large and supersymmetric particles do not appear in the
Higgs decay products, the full parameter space in the conventional ($m_A,
\tan \beta$) plane should be covered. The $5 \sigma$ discovery contour
curves
are shown in Fig.~\ref{fig:susycontour} for individual channels and for
integrated luminosities of 300~$\Ifb$. In addition to the
channels discussed for the Standard Model case, the MSSM Higgs search
relies heavily on the $H/A \rightarrow \tau \tau, \mu \mu$ channel,
and on the charged Higgs decays. Over a large fraction of the
parameter space more than one Higgs boson and/or more than one decay mode
would be accessible. 
%Over a large part, the experiments would be able to
%distinguish between the Standard Model and the MSSM models. 

The interplay between SUSY particles and the Higgs sector has also been
addressed in the ATLAS studies. SUSY scenarios have an impact on the
discovery potential through the opening of Higgs-boson decays to SUSY
particles (mostly for H and A) and through the presence of SUSY particles
in loops.

%%%%%%%%%%%%%%%%
%Scenarios in which SUSY particles are light and
%appear as Higgs decay products have been studied in the framework of
%SUGRA
%models. The discovery potential of the lightest neutral Higgs $h$ would 
%not be significantly different from what is obtained in the heavy SUSY
%scenario, since within the model, given present experimental constraints,
%the decay of the $h$ to the lightest SUSY particles is kinematically
%forbidden. Moreover, over a large fraction of the 
%%%%%%%%%%%%%%%%%%%%

\section{Determination of the Standard Model Higgs boson parameters}
Assuming that a Standard Model Higgs boson will have been discovered at
the
LHC, the potential for the precision measurement of the Higgs parameters
(mass, width, production rates, branching ratios) is discussed in this
section. Such measurements should give further insights into the
electroweak symmetry-breaking mechanism and into the way the Higgs couples
to fermions and bosons, and, in some cases, should allow a distinction
between a Standard Model and a MSSM Higgs boson. 
\subsection{Measurement of the Higgs boson mass and width}
\begin{figure}
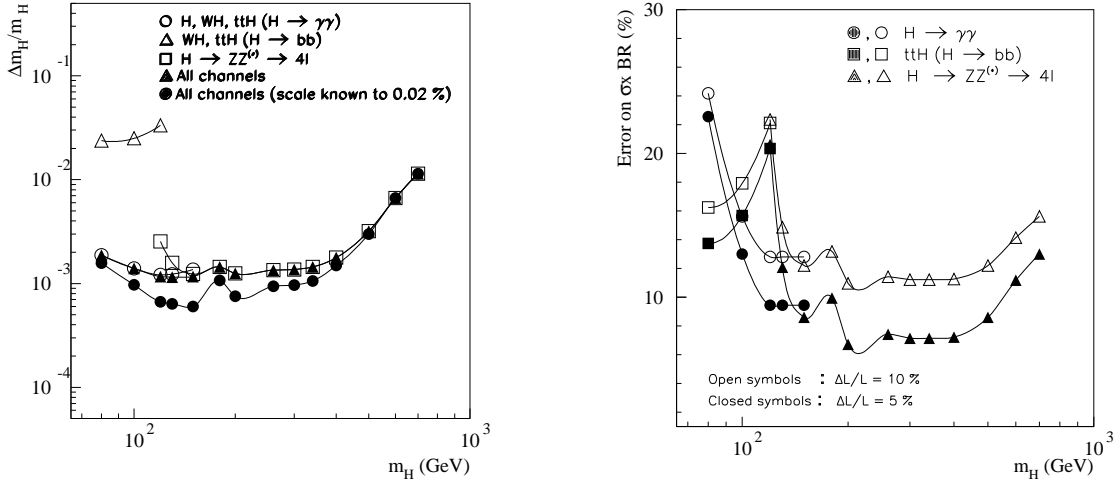

\begin{center}
\psfig{figure=tdrdmh.epsi,height=2.5in} ~~~~~~~~~~~
\psfig{figure=tdrbr.epsi,height=2.5in} 
\end{center}
\caption{Left: relative experimental precision $\Delta m_H / m_H$ on the
measured Higgs
boson mass as a function of $m_H$ for different Higgs decay
channels. Right: relative precision on the measurements of the Higgs boson
rate for various channels as a function of $m_H$. An integrated luminosity
of 300~$\Ifb$ is assumed in both plots. 
 \label{fig:massres}} \end{figure}

The ultimate experimental precision with which the Higgs boson mass will
be measured in shown in Fig.~\ref{fig:massres} for the ATLAS detector and
assuming 300~$\Ifb$ of integrated luminosity. The results obtained in the
various decay channels as well as the combination of all channels are
given. The quoted precision includes the statistical error in the
determination of the peak position, coming from both the limited number of
signal events, the error on the background subtraction,
%(the background
%is assumed to be flat under the peak)
 and the systematic error on the
absolute energy scale. The latter is conservatively assumed to be 0.1\%
for decay
channels which contain leptons or photons 
and 1\% for decay
channels containing jets.
Fig.~\ref{fig:massres} indicates that  the Higgs mass can be
measured with a precision of 0.1\% up to masses of 400~GeV. For larger
masses the precision deteriorates because the Higgs boson width becomes
large and the statistical error increases. 
%However even for masses as
%large as 700~GeV the Higgs boson mass can be measured with an accuracy of
%1\%. 
%The precision of the measurement is determined by the four-lepton and
%two-photon channels, whereas the $H \rightarrow \Bbbar$ channel
%contributes very little. This is due to both the larger systemaic error
%on
%the absolute energy scale compared to the lepton/photon scale and the
%larger statistical error because of worse mass resolution ($\approx$
%15~GeV for $H \rightarrow \Bbbar$, compared to $\approx$ 1.5~GeV for $H
%\rightarrow \gamma  \gamma$ and $H \rightarrow ZZ^{(*)} \rightarrow 4l$).
%%
%%
%
%\subsection{Measurement of the Higgs boson width}

The Higgs boson width can be experimentally obtained from a measurement of
the width of the reconstructed Higgs peak, after unfolding the
contribution of the detector resolution. This direct measurement is only
possible for Higgs boson masses larger than 200~GeV, above which the
intrinsic width of the resonance becomes comparable or larger than the
experimental mass resolution. This is the mass region covered mainly by
the $H \rightarrow ZZ^{(*)} \rightarrow 4l$ decays. Over the range $300 <
m_H < 700$~GeV, the precision of the measurement is approximately constant
and of the order of 6\%. 
%This result includes the statistical uncertainty
%coming from the number of signal events and the systematic uncertainty
%coming from the measurement of the peak width and the knowledge of the
%detector energy and momentum resolution. In both cases, the systematic
%error is dominated by the the uncertainty on the radiative decays.
%%
%%
\subsection{Measurement of the Higgs boson rate}
The measurement of the Higgs boson rate in a given decay channel provides
a measurement of the production cross-section times the decay branching
ratio for that channel. Such measurement in some cases would help to
disentangle between Standard Model and MSSM Higgs scenarios. 

The statistical error on such measurements is expected to be smaller than
10\% over the mass region 120--600~GeV using the $\gamma \gamma$,
$\Bbbar$ and $4l$ final states. The main systematic error comes from the
knowledge of the luminosity and background subtraction.
% and two values have been considered for the
%luminosity uncertainty: 5\% (a somewhat ambitious goal) and 10\% (a more
%conservative estimate). An additional systematic error of 10\% have been
%included to take into account the uncertainty on the background
%subtraction for channels where the background is not completely flat
%under
%the peak (e.g. $\Ttbar H$). 
Fig.~\ref{fig:massres} shows the expected
experimental uncertainty on the Higgs boson rates, for various production
and decay channels and for two assumptions on the luminosity
uncertainty.
 By performing these measurements for several channels, one can obtain
several
constraints on the Higgs boson couplings.
% to fermions and bosons, which
%can
%be used to test the theory.

\section{Conclusions}
The  ATLAS and CMS experiments at the LHC should discover a Higgs boson
(with significance above 5) over the full allowed mass region from the
limit set by previous machines up to the theoretical bound of 1~TeV, after
less than two years of operation. 

Different channels cover different mass regions, but over most of the
mass range more than one channel should be observed, thus giving
robustness to the discovery and hints to understand the nature of the
signal.

If a Standard Model Higgs boson will be observed at the LHC, then ATLAS
and CMS should be able to measure its mass with a precision of
$\approx$0.1\% and to study its coupling to SM particles with a  
$\approx$10\% precision.

%\section*{Acknowledgments}
%This is where one places acknowledgments for funding bodies etc.
%Note that there are no section numbers for the Acknowledgments, Appendix
%or References.

\end{document}

%% file: my_definitions.tex
\newcommand {\Qqbar}    {q\bar{q}}
\newcommand {\Ttbar}    {t\bar{t}}
\newcommand {\Bbbar}    {b\bar{b}}

\newcommand {\Ifb}      {{{\rm fb}^{-1}}}